# Kinetic Scale Slow Solar Wind Turbulence in the Inner Heliosphere: Co-existence of Kinetic Alfvén Waves and Alfvén Ion Cyclotron Waves

S. Y. Huang[1], J. Zhang[1], F. Sahraoui[2], J. S. He[3], Z. G. Yuan[1], N. Andrés[4,5], L. Z. Hadid[2], X. H. Deng[6], K. Jiang[1], L. Yu[1], Q. Y. Xiong[1], Y. Y. Wei[1], S. B. Xu[1], S. D. Bale[7], and J. C. Kasper[8]

[1]School of Electronic Information, Wuhan University, Wuhan, 430072, China

[2]Laboratoire de Physique des Plasmas, CNRS-Ecole Polytechnique-Sorbonne Université-Univ, Paris-Saclay-Observatoire de Paris-Meudon, Palaiseau, F-91128, France

[3]School of Earth and Space Sciences, Peking University, Beijing, 100871, China

[4]Instituto de Astronomía y Física del Espacio, CONICET-UBA, Ciudad Universitaria, 1428, Buenos Aires, Argentina

[5]Departamento de Física, Facultad de Ciencias Exactas y Naturales, UBA, Ciudad Universitaria, 1428, Buenos Aires, Argentina

[6]Insititute of Space Science and Technology, Nanchang University, Nanchang, 330031，China

[7]Space Sciences Laboratory and Physics Department, University of California, Berkeley, CA 94720-7450, USA

[8]Department of Climate and Space Sciences and Engineering, University of Michigan, Ann Arbor, MI 48109, USA

## Abstract

The nature of the plasma wave modes around the ion kinetic scales in highly Alfvénic slow solar wind turbulence is investigated using data from the NASA's Parker Solar Probe taken in the inner heliosphere, at 0.18 Astronomical Unit (AU) from the sun. The joint distribution of the normalized reduced magnetic helicity $\sigma_m$ ($\theta_{RB}$, $\tau$) is obtained, where $\theta_{RB}$ is the angle between the local mean magnetic field and the radial direction and $\tau$ is the temporal scale. Two populations around ion scales are identified: the first population has $\sigma_m$ ($\theta_{RB}$, $\tau$) <0 for frequencies (in the spacecraft frame) ranging from 2.1 to 26 Hz for 60º < $\theta_{RB}$ < 130º, corresponding to kinetic Alfvén waves (KAWs), and the second population has $\sigma_m$ ($\theta_{RB}$, $\tau$) >0 in the frequency range [1.4, 4.9] Hz for $\theta_{RB}$ > 150º, corresponding to Alfvén ion Cyclotron Waves (ACWs). This demonstrates for the first time the co-existence of KAWs and ACWs in the slow solar wind in the inner heliosphere, which contrasts with previous observations in the slow solar wind at 1 AU. This discrepancy between 0.18 and 1 AU could be explained, either by i) a dissipation of ACWs via cyclotron resonance during their outward journey, or by ii) the high Alfvénicity of the slow solar wind at 0.18AU that may be favorable for the excitation of ACWs.

# 1. Introduction

Turbulence is thought to contribute significantly to particle heating in various space astrophysical plasmas (e.g., Tu & Marsch, 1995; Bruno & Carbone, 2013; Goldstein et al., 2015; Huang et al., 2017b; Andrés et al., 2019; Sahraoui et al., 2020). Because of the collisionless nature of the near-Earth space plasmas (e.g., the solar wind and the magnetosheath), energy dissipation into particle heating is thought to occur via a variety of processes that include resonant wave-particle interactions, e.g. Landau damping (e.g., Sahraoui et al., 2009; Chen et al., 2019), cyclotron damping (e.g., He et al., 2015; Woodham et al., 2018), stochastic heating (Chandran et al., 2010; Bourouaine & Chandran, 2013; Bowen et al., 2020b), and intermittent heating within coherent structures (e.g., Retinò et al., 2007; Sundkvist et al., 2007; Wang et al., 2013; Chasapis et al., 2015; Zhang et al., 2015; Huang et al., 2017a, 2017c, 2018). Identifying the dissipation processes at work requires unraveling the plasma wave modes dominating the cascade, in particular at the sub-ion scales.

At 1 Astronomical Unit (AU), the solar wind is generally categorized according to its velocity: fast solar wind ($V_f \geq 500$ km/s) and slow solar wind ($V_f < 500$ km/s). Intensive research work has been dedicated to identifying the nature of the wave modes in the solar wind at 1AU. Kinetic Alfvén waves (KAWs) are characterized by a quasi-perpendicular wave vector ($k_\parallel \ll k_\perp$) and a right-handed polarization (e.g., Howes et al., 2010; Sahraoui et al., 2012; Zhao et al., 2013, 2016). They have been identified at kinetic scales in the fast solar wind turbulence in several studies that used in-situ data. For instance, Leamon et al. (1998) have analyzed solar wind magnetic fluctuations in the sub-ion (dissipation) range and found that the 2D component is consistent with KAWs propagating at large angles respect to the background magnetic field. Bale et al. (2005) and Sahraoui et al. (2009) have established the wave dispersion based on the electric and magnetic field spectra, and found the wave dispersion around ion scales are consistent with KAWs. A subsequent work by Sahraoui et al. (2010) provided a direct evidence of the dominance of the of KAW mode at the sub-ion scales using the multi-point measurement technique, namely the *k*-filtering technique, on the

Cluster data, which allowed them to obtain the 3-dimentional dispersion relation that agreed well with the theoretical predictions for KAWs. Chen et al. (2012) used the spectral indices of the magnetic field and electron density at the kinetic scales to identify the nature of solar wind turbulence, and were found to be consistent with the numerical simulation results of KAW turbulence (e.g., Howes et al., 2011). He et al. (2011, 2012a 2012b, 2015) used the reduced (fluctuating) magnetic helicity (Matthaeus & Goldstein, 1982), estimated as a function of the angle $\theta_{BV}$ (or $\theta_{RB}$) between the direction of the local mean magnetic field and the solar wind velocity (or the radial direction), and showed that a major population of magnetic fluctuations in the dissipation range has quasi-perpendicular angles (relative to the local mean magnetic field) and right-handed polarization, consistent with quasi-perpendicular KAWs. Subsequent studies by Podesta & Gary (2011), Podesta & Tenbarge (2012), Podesta (2013), and Bruno & Telloni (2015) confirmed those findings.

On the other hand, a minor component of the fast and slow solar wind turbulence was found to be left-handed polarization and has parallel propagation. This mode is known as the Alfvén ion cyclotron waves (ACWs), electromagnetic ion cyclotron (EMIC) waves or ion cyclotron waves (ICWs). Jian et al. (2009, 2010) have shown sporadic observations of ACWs with short time intervals and found that the ACWs have a preference for radial field alignment. He et al. (2011, 2012a, 2015) showed the existence of ACWs in the fast solar wind when the velocity and magnetic field were quasi-aligned ($\theta_{BV} < 30°$) (See also Telloni et al. (2019)). Recently, Bowen et al. (2020a) have used the observations from Parker Solar Probe (PSP) mission to investigate ion-scale electromagnetic waves in the inner heliosphere, and revealed that 30-50% of radial field intervals have parallel/anti-parallel propagation and circularly polarized waves.

The slow solar wind has low amplitude magnetic fluctuations compared to the fast solar wind at 1 AU (e.g., Dasso et al., 2005; Bruno et al., 2014). D'Amicis & Bruno (2015) have shown two different kinds of slow solar wind: one coming from coronal streams or active regions with low Alfvénicity, and the other one from the boundary of coronal holes with highly Alfvénicity. Further, D'Amicis et al. (2019) observed Alfvénic slow wind at 1 AU during a maximum of the solar activity, and found it be similar to fast solar wind than to typical non-Alfvénic slow wind. Thus, they suggested that the Alfvénic slow wind and fast solar wind probably have a similar solar origin. On the other hand, it is found that, at solar maximum, 34% of the slow solar wind streams ($V_f$ < 450 km s$^{-1}$) with quiet-sun as their source region are featured with high Alfvénicity ($|\sigma_c|$ > 0.7) (Wang et al., 2019). Accordingly, Wang et al. (2019) suggested that the slow solar wind streams from the quiet-Sun region, like their counterparts from coronal hole region, can directly flow outward along the open field lines. A similar scenario for the origin of solar wind as emerging from quiet sun region had already been proposed by He et al. (2007). Recently, Alfvénic slow wind has also been observed in the inner heliosphere at 0.3 AU during a minimum of solar activity using Helios data (e.g., Stansby et al., 2019, 2020; Perrone et al., 2020). Moreover, Bale et al. (2019) have demonstrated that the slow Alfvénic solar wind from 0.17 to 0.25 AU measured during PSP Encounter 1 emerges from a small equatorial coronal hole. Moreover, Bale et al. (2019) have shown the co-existence of electron and ion micro-instabilities in this slow solar wind.

Despite a big progress in understanding the slow solar wind physics, the nature of the wave modes dominating the turbulence cascade is still an unsettled problem, especially in the inner heliosphere. In the present study, using the NASA's Parker Solar Probe observations at 0.18 AU, we bring new insight into this problem. Our main finding is the co-existence of kinetic Alfvén waves and Alfvén ion cyclotron waves for highly Alfvénic slow solar winds.

## 2. Data Analysis and Results

In the present study, the solar wind proton moments were measured by Solar Wind Electron, Alpha, Proton (SWEAP) experiment on PSP with sampling frequencies between 1 Sa/cycle and 4 Sa/cycle, where 1 cycle is approximately equal to 0.873 s (Kasper et al., 2016; Case et al., 2020). The magnetic field data were measured at the sampling frequency of 256 Sa/cycle (~293 samples/sec) by the FIELDS flux-gate magnetometer (Bale et al., 2016) for the Encounter mode. All vector data are presented in the radial tangential normal (RTN) coordinate system.

The normalized (fluctuating) reduced magnetic helicity ($\sigma_m$) is useful to diagnose polarization characteristics of solar wind turbulence (Matthaeus & Goldstein, 1982), which can be linked to the classical wave polarization of the fluctuations (see, e.g. Howes & Quataret, 2010; Meyrand & Galtier, 2012; Klein et al., 2014). Here we used the method developed in He et al. (2011, 2015), where the $\sigma_m$ spectra are estimated as function of angle $\theta_{RB}$ to account for the local (in time) variations of the mean magnetic field.

A windowed Fourier transform is performed for the magnetic field to obtain a time-frequency decomposition of the power spectral densities (PSD ($t,\tau$)) of the magnetic field and the normalized reduced magnetic helicity $\sigma_m$ ($t, \tau$) ranging from -1 to +1, where $t$ and $\tau$ are the measurement time and temporal scale, respectively. To fit the spectral densities at different times and temporal scales, the time-frequency spectral indices (or slope ($t, \tau$)) can be obtained. The local mean magnetic field $\boldsymbol{B}_0(t, \tau)$ is calculated by the Equation (22) from Podesta (2009), then one can obtain the angle $\theta_{RB}$ ($t, \tau$) between the radial direction and the local mean magnetic field (ranging from 0 to 180º).

Figure 1 shows the PSP spacecraft observations on 6 November 2018 in the perihelion at 0.18 AU. The average plasma parameters are: $|B| \sim 89$ nT, the proton density $n_p \sim 315$ cm$^{-3}$, and the proton temperature $T_p \sim 36$ eV, yielding the Alfvén speed $V_A \sim 109$ km/s, the proton inertial length $d_i \sim 13$ km, and the proton gyro-radius $\rho_i \sim 9.7$ km. One can see that the magnetic field is well correlated with the proton velocity (correlation coefficient > 0.82, in Figure 1a-1c), indicating highly Alfvénic fluctuations in this time interval (Kasper et al., 2019). The mean $V_R$ is about 360 km/s, which is smaller than 500 km/s, indicating that PSP encountered the slow solar wind. $B_R$ is mostly negative (Figure 1a), implying that PSP was in an inward magnetic sector. The magnetic field has large amplitude fluctuations compared to typical slow solar wind (e.g., Bruno et al., 2014). The PSD of the magnetic field shows a scaling close to the Kolmogorov spectrum in the lowest frequency range (with some fluctuations due the limited small size window used to fit the local slopes), before steepening to ~ -4 above the spectral break, then flattening for frequencies > 20 Hz due to reaching the noise floor of the instrument (Figure 1d). The red and blue bars at higher frequencies are due to interference from the spacecraft (reaction wheels signal). The magnetic helicity $\sigma_m$ $(t, \tau)$ is illustrated in Figure 1f. It varies randomly at low frequency, but shows a coherent pattern at high frequencies: often positive around 3 Hz and permanently negative around 7 Hz, corresponding to the steep spectra in Figure 1d-e. The angles between the radial direction and the local mean magnetic field direction $\theta_{RB}$ is shown in Figure 1g. It is found that the angle $\theta_{RB}$ varies in time over the range from 40º-180º, while it does not change much in frequency, which indicates that the large scale magnetic field dictates the behavior of the radial angle at all scales. This does not contradict the observations that the fluctuations, and thus the vector orientation, are random (i.e. noise) at $f$ >20Hz so long as the amplitudes of the high frequency fluctuations are much smaller compared to the static (large scale) field.

Figure 2 displays the time-averaged (a) PSD of the magnetic field and (b) the magnetic helicity $\sigma_m$ as a function of the frequency. One can identify two distinct ranges from the PSD: i) a Kolmogorov-like inertial range from 0.04 Hz to ~ 1 Hz, i.e., at the MHD scales; ii) a transition range around ion scales with a steep slope (up to -4.24). The time-averaged $\sigma_m$ has very small negative values (close to zero) at MHD scales, consistent with previous observations (e.g., Goldstein et al., 1994; He et al., 2011; Podesta, 2013). It is interesting that $\sigma_m$ changes its polarity at 1.4 Hz, and then become negative above 4.2 Hz. The sign change of $\sigma_m$ contrasts with previous observations of net (non-zero) right-handed polarity, which has been explained by the damping of left-handed fluctuations by cyclotron resonance at ion scales while whistler or KAW waves survive and carry the turbulent cascade at smaller scales (e.g., Goldstein et al., 1994; Howes & Quataret, 2010; He et al., 2011, Podesta, 2013; Klein et al., 2014; Woodham et al., 2019).

Based on the angle $\theta_{RB}$ ($t,\tau$) and the magnetic helicity $\sigma_m$ ($t,\tau$), we constructed the joint distribution $\sigma_m$ ($\theta_{RB},\tau$) in Figure 3a. It can be clearly seen that there are two populations in $\sigma_m(\theta_{RB},\tau)$: the first population has negative magnetic helicity corresponding to $\theta_{RB} \in$ [60º 130º] and frequencies $\in$ [2.1, 26] Hz, the second population has positive magnetic helicity corresponding $\theta_{RB} \in$ [150º, 180º] and frequencies $\in$ [1.4, 4.9] Hz. For an inward-oriented background magnetic field (namely inward magnetic sector with $B_R$ <0), a left-handed polarized wave mode has positive magnetic helicity, while a right-handed polarized wave mode has negative magnetic helicity (e.g., He et al., 2011; Bruno & Telloni, 2015). The magnetic fluctuations with very small or large $\theta_{RB}$ (i.e., close to 0º or 180º) correspond to waves propagating quasi-parallel or quasi-anti-parallel to the mean magnetic field; while the magnetic fluctuations with the intermediate $\theta_{RB}$ (i.e., close to 90º) correspond to waves propagating quasi-perpendicular to the mean magnetic field (e.g., He et al., 2011, 2015). Therefore, the magnetic fluctuations with positive helicity at low frequency and around 180° can be identified as quasi-parallel left-handed ACWs, while the magnetic fluctuations with negative helicity at high frequency and around 90° are likely to be quasi-perpendicular right-handed KAWs.

Finally, the magnetic trace power spectra and magnetic helicity $\sigma_m$ for two angular ranges, i.e., 80º < $\theta_{RB}$ < 100º and 150º < $\theta_{RB}$ < 180º, are shown in Figure 3b and 3c. The PSD for quasi-perpendicular angles (blue line) is slightly higher than the one for quasi-parallel angles (red line). The PSD in the perpendicular direction at low frequency (MHD scales) has a scaling close to the Kolmogorov spectrum, while the PSD in the (anti) parallel direction is steeper ($f^{-1.8}$). Both show a spectral break around 1.7 Hz. At higher frequencies, the PSD in the (anti) parallel direction shows a steeper transition range with a slope close to -5, in comparison to -3.73 of in the perpendicular direction. We note a slight pump between 1.7 to 4.6 Hz in the PSD for quasi (anti-) parallel direction, corresponding to the frequency range where positive $\sigma_m$ is observed, which might indicate that the bump is caused by the ACWs (Figure 3c).

## 3. Discussion and Conclusions

We investigated the nature of the kinetic wave modes in the slow solar wind using data from the NASA's PSP spacecraft at 0.18 AU. To the best of our knowledge, this is the first time that co-existing of KAWs and ACWs in the slow wind in the inner heliosphere is revealed.

The co-existence of KAWs and ACWs in the slow solar wind is quite similar to previous observations in the fast solar wind at 1 AU (e.g., He et al., 2011, 2015; Podesta, 2012, 2013), but inconsistent with other observations in the slow solar wind at 1 AU (Burno & Telloni, 2015) which showed the disappearance of the ion-cyclotron signature of the magnetic helicity followed by a more gradual disappearance (or weakness) of KAWs with the decrease of solar wind speed. Those results were later confirmed by Woodham et al. (2018). If one assumes that the ACWs appear in the slow solar wind at 0.18 AU but disappear at 1 AU, then that would imply that the ACWs heat the plasma protons via cyclotron resonance during the outbound traveling from the inner heliosphere to the outer heliosphere, until they vanish at 1 AU. However, the difference between the observation of ACWs at 0.18 AU and 1AU can be due to difference in the Alfvénicity of fluctuations: the slow solar wind in Burno and Telloni (2015) has low Alfvénic fluctuations, but in our case the slow solar wind is highly Alfvénic. Another possible explanation may come from the possible generation mechanism of the ACWs: because of the existence of a drift of alpha particles with respect to the protons, the proton temperature anisotropy instability that operates when $T_{p\perp}/T_{p\|} > 1$ preferentially generates outward propagating ion-cyclotron (Podesta & Gary, 2011; Woodham et al., 2019). This condition might be met more preferentially in the inner heliosphere than at 1AU. A future study based on the radial the evolution of both the proton parallel and perpendicular temperature should help deciding between the different possible explanations.

Another open question is how important are the ACWs in the overall dynamics of the turbulence cascade at the sub-ion scale. We estimated that about 50% of time the ACWs were observed in our (1 day) data. However, this should be balanced by the level of power that is carried by the parallel component of the fluctuations (see Fig. 3b). The integral power densities of the frequency range corresponding to the positive magnetic helicity for both KAWs and ACWs are estimated. It is found that the integral power of KAWs ($dB^2$(KAW) ~ 10.98 nT$^2$) is higher than the power of ACWs ($dB^2$(ACW) ~ 4.52 nT$^2$, i.e., ~ 41% $dB^2$(KAW)), implying that the ACWs play non-negligible role in the slow solar wind.

One of the important first results from PSP is the ubiquity of the so-called 'switchbacks', which are probably large Alfvén waves (Kasper et al., 2019). Although the switchback does not affect directly the calculation of the magnetic helicity (i.e., it has no dependence on the sign of $B_R$ –see, e.g., Eq. 1 in He et al. (2011)), it may however affect the determination of the wave polarization from the magnetic helicity, where the sign of $B_R$ enters into play (Howes et al., 2010; He et al., 2011). Therefore, if $B_R$ changes sign (or equivalently $\theta_{RB}$ changes from <90° to >90° or vice versa) on short time scales corresponding to those where the change of polarity occurs then this may affect the results of the study. However, as can be seen in Fig. 1g and discussed above, the angle $\theta_{RB}$ does not change significantly in the frequency range ~0.1-20 Hz in which we reported the change of polarity, but it may change on larger time scales. This observation is in agreement with those of Dudok to Wit et al. (2020) who analyzed a larger data set (that included our time interval) and found that switchback affect larger scales that belong to the inertial range (or even to the 1/$f$ range). A similar finding is reported concerning the cross helicity (McManus et al., 2020). The fact that we observe two wave modes with distinct polarities at high frequency, which extend over relatively broad frequency bands are argument that would exclude a possible role of the switchbacks.

A final key point that is worth discussing here is a possible ambiguity in the interpretation of the spectral slopes observed in Figure 3b for the (anti-)parallel and perpendicular spectra that might stem from the sampling direction of the fluctuation due to the flow motion (w.r.t. the spacecraft). Indeed, the critical balance (CB) conjecture predicts an anisotropic scaling of the for Alfvénic turbulence: $l_\parallel \propto l_\perp^{2/3}$ at MHD scales and $l_\parallel \propto l_\perp^{1/3}$ at the sub-ion scales (Goldreich & Sridhar, 1994; Schekochihin et al., 2009). These scaling results in reduced spectra for the magnetic fluctuations given by: $\delta B^2 \propto k_\parallel^{-2}$ at MHD scales and $\delta B^2 \propto k_\parallel^{-5}$ at the sub-ion scales. Using the Taylor hypothesis and considering that $\theta_{RB} \sim \theta_{VB}$, the parallel spectrum of Figure 3b translates into $\delta B^2 \propto k_\parallel^{-1.8}$ and $\delta B^2 \propto k_\parallel^{-4.93}$ for MHD and sub-ion scales, respectively. This estimation is a good accord with the CB prediction (see Horbury et al. (2008) and Podesta (2009) for a similar conclusion regarding MHD scale turbulence in the fast solar wind). This would mean that the spectral features of Figure 3b can be fully explained by a simple sampling effect of Alfvénic and KAW turbulence and no need for evoking the ACWs. However, the sole presence of KAW cannot explain the change of polarity observed in $\sigma_m$ around the ion scales. A further insight can be gained by examining the range of parallel scales involved in Figure 3b-3c. The ACWs seem to be observed within the frequency range

[1.4, 4.9] Hz. Using the Taylor hypothesis this translates into the scale range $k_\parallel d_i \in [0.4, 1.2]$, given $V_f \sim V_R \sim 360$ km/s and $d_i \sim 13$ km. This range of scales corresponds to those where the cyclotron damping is expected to be effective (Gary & Borovsky, 2004). At these scales, the KAW turbulence is expected to have $k_\parallel \rho_i \ll 1$ at $k_\perp \rho_i \sim 1$ (Sahraoui et al., 2010). This argument supports the scenario of ACW to explain the polarity near $k_\parallel d_i \sim 1$.

On the other hand, the scaling of PSD in the perpendicular direction $\delta B^2 \propto k_\perp^{-1.56}$ at MHD scales and $\delta B^2 \propto k_\perp^{-3.73}$ at the sub-ion scales agree with the CB prediction at MHD scales ($k_\perp^{-5/3}$) but is steeper than that in the sub-ion scales ($k_\perp^{-7/3}$). This steeping might be caused by a dissipation of part of the KAWs into ion heating via Landau damping as suggested in Sahraoui et al. (2010) and shown in numerical simulations of Howes et al. (2011) and Kobayashi et al. (2017). Note that theories and numerical simulations of incompressible Hall-MHD turbulence predicts a scaling $k^{-11/3}$ for the left-handed component (ACW) and $k^{-7/3}$ for right-handed component (KAW) of the turbulence (Meyrand & Galtier, 2012), although that model does not capture all aspects of the ACW and KAW modes, which are inherently kinetic in nature.

**Acknowledgement**

This work was supported by the National Natural Science Foundation of China (41674161, 41874191, 41925018), Young Elite Scientists Sponsorship Program by CAST (2017QNRC001), and the National Youth Talent Support Program. We thank the entire PSP team and instrument leads for data access and support. The SWEAP and FIELDS investigation and this publication are supported by the PSP mission under NASA contract NNN06AA01C. PSP data is publicly available from the NASA's Space Physics Data Facility (SPDF) at https://spdf.gsfc.nasa.gov/pub/data/psp/.

Zhao, J. S., Y. M. Voitenko, D. J. Wu, and M. Y. Yu, 2016, JGR, 121, 5–18

Zhao, J. S., Wu, D. J., Lu, J. Y., ApJ, 2013, 67(2): 0~109

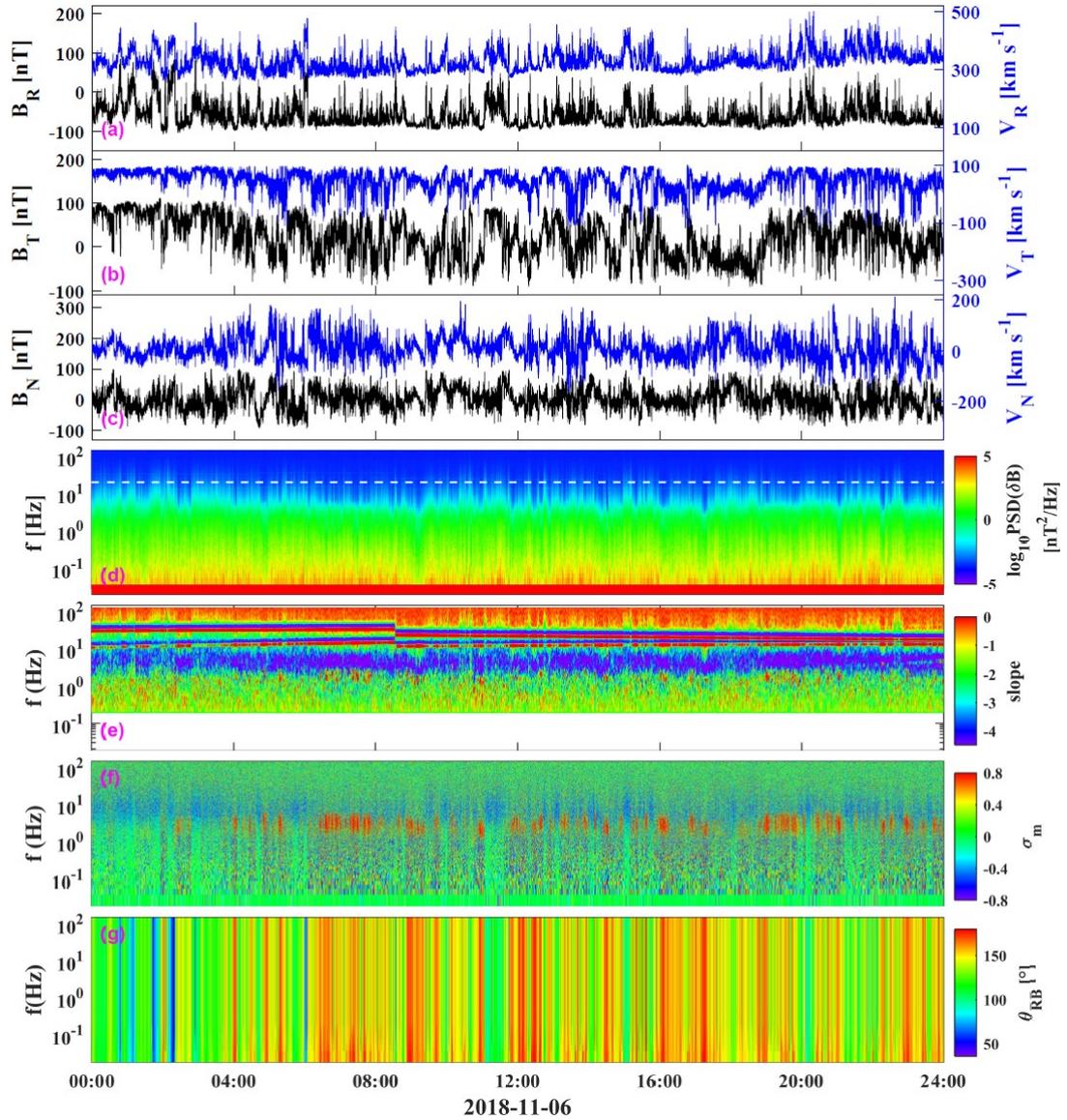

Figure 1. The overview observations of slow solar wind in 06 November 2018. (a-c) three magnetic field component $B_R$, $B_T$, $B_N$ and three proton velocity component $V_R$, $V_T$, $V_N$, respectively; (d) power spectral density of magnetic field; (e) the spectral slopes; (f) the normalized reduced magnetic helicity $\sigma_m$; (g) the angle $\theta_{RB}$ between the local mean magnetic field and the radial direction.

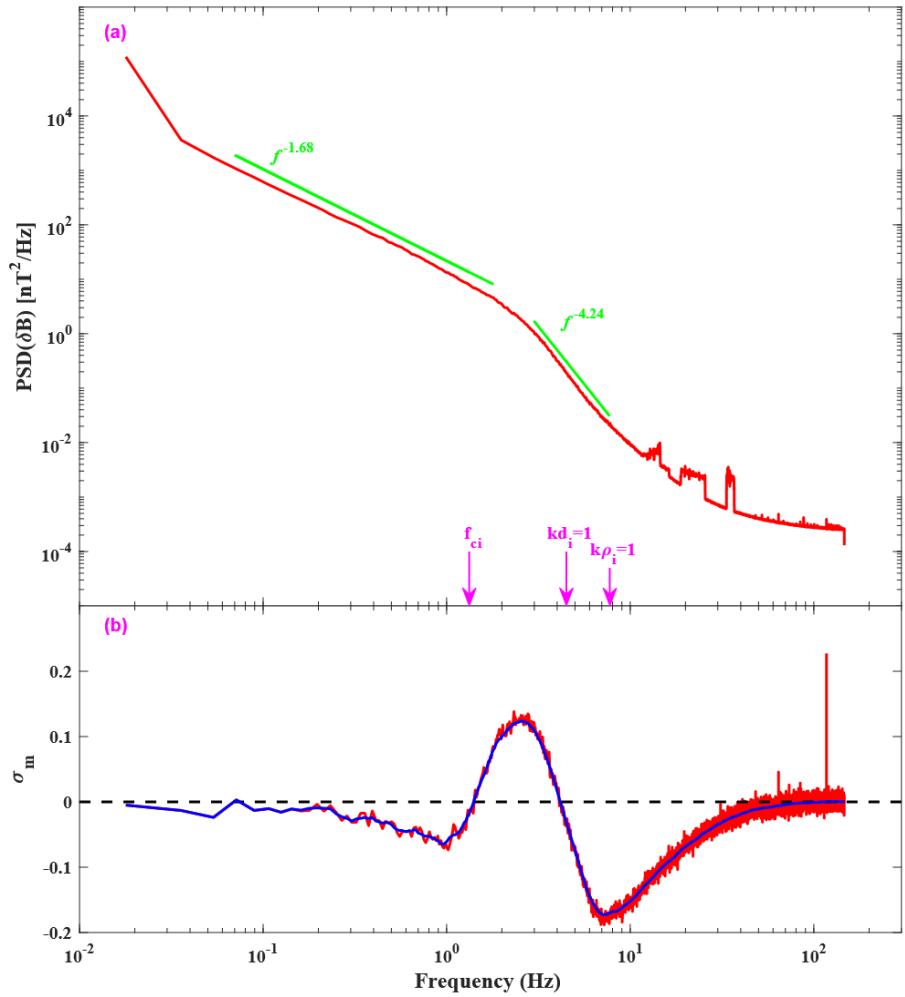

Figure 2. The one-day time-averaged (a) PSD of magnetic field and (b) the magnetic helicity $\sigma_m$ (the red and blue curves present the origin and smoothed $\sigma_m$). The three vertical arrows from left to right correspond to the proton cyclotron frequency, the Doppler shifted frequency of the proton inertial length and proton gyro-radius, respectively. Discrete spectral features above 10 Hz is noise from the spacecraft momentum wheels.

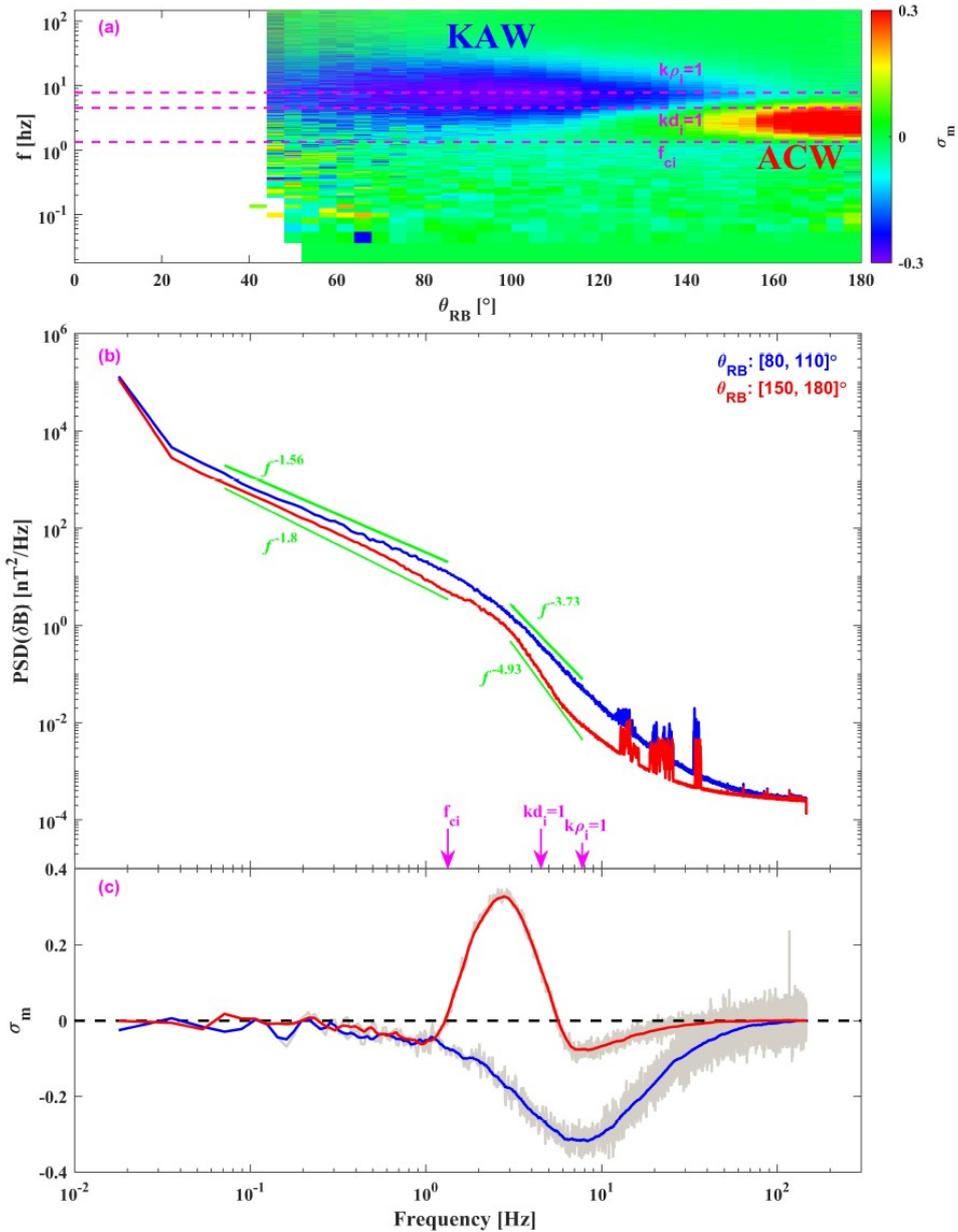

Figure 3. (a) The statistical joint distribution $\sigma_m$ ($\theta_{RB}$, $\tau$) of the whole day, showing the signatures of quasi-parallel left-handed polarization ACWs ($\sigma_m > 0$ at $\theta_{RB} > 150º$) and quasi-perpendicular right-handed polarization KAWs ($\sigma_m < 0$ at $60º < \theta_{RB} < 130º$); (b) Magnetic trace power spectra and (c) magnetic helicity $\sigma_m$ for two angular ranges ($80º < \theta_{RB} < 100º$ in blue and $150º < \theta_{RB} < 180º$ in red). Discrete spectral features above 10 Hz is noise from the spacecraft momentum wheels. The grey curves present the origin $\sigma_m$, and the red and blue curves present the smoothed $\sigma_m$ in (c).